\documentclass[12pt,preprint]{aastex}

\begin{document}

\shorttitle{Millimeter Structure in AU Mic}
\shortauthors{MacGregor et al.}

\slugcomment{accepted by ApJ Letters: November 17, 2012}

\title{
Millimeter Emission Structure in the first ALMA
Image of the AU Mic Debris Disk
}

\author{
Meredith A. MacGregor\altaffilmark{1},
David J. Wilner\altaffilmark{1},
Katherine A. Rosenfeld\altaffilmark{1},
Sean M. Andrews\altaffilmark{1}, \\
Brenda Matthews\altaffilmark{2},
A. Meredith Hughes\altaffilmark{3},
Mark Booth\altaffilmark{2,4},
Eugene Chiang\altaffilmark{3}, \\
James R. Graham\altaffilmark{3,5},
Paul Kalas\altaffilmark{3,6},
Grant Kennedy\altaffilmark{7},
Bruce Sibthorpe\altaffilmark{8}
}
\altaffiltext{1}{Harvard-Smithsonian Center for Astrophysics, 
  60 Garden Street, Cambridge, MA 02138, USA}
\altaffiltext{2}{Herzberg Institute of Astrophysics, 
  5072 West Saanich Road, Victoria, BC V9E 2E7, Canada}
\altaffiltext{3}{Department of Astronomy, 601 Campbell Hall, 
  University of California, Berkeley, CA 94720, USA} 
\altaffiltext{4}{Deptarment of Physics \& Astronomy, University of
  Victoria, 3800 Finnerty Rd., Victoria, BC, V8P 5C2, Canada}
\altaffiltext{5}{Dunlap Institute for Astronomy \& Astrophysics, 
  University of Toronto, Toronto, ON, Canada}
\altaffiltext{6}{SETI Institute, 
  189 Bernardo Ave., Mountain View, CA 94043}
\altaffiltext{7}{Institute of Astronomy, University of Cambridge, 
  Madingley Road, Cambridge CB3 0HA, UK}
\altaffiltext{8}{SRON Netherlands Institute for Space Research, 
  NL-9747 AD Groningen, The Netherlands}

\begin{abstract}
We present 1.3\,millimeter ALMA Cycle~0 observations of the edge-on debris disk 
around the nearby, $\sim$10\,Myr-old, M-type star AU~Mic.  These observations 
obtain $0\farcs6$ (6\,AU) resolution and reveal two distinct emission 
components: (1) the previously known dust belt that extends to a radius of
$40$\,AU, and (2) a newly recognized central peak that remains unresolved.  The 
cold dust belt of mass $\sim1~M_{\rm Moon}$ is resolved in the radial direction 
with a rising emission profile that peaks sharply at the location of the outer 
edge of the ``birth ring'' of planetesimals hypothesized to explain the 
midplane scattered light gradients.  No significant asymmetries are discerned 
in the structure or position of this dust belt.  The central peak identified in 
the ALMA image is $\sim6$ times brighter than the stellar photosphere, which 
indicates an additional emission process in the inner regions of the system.  
Emission from a stellar corona or activity may contribute, but the observations 
show no signs of temporal variations characteristic of radio-wave flares.  We 
suggest that this central component may be dominated by dust emission from an 
inner planetesimal belt of mass $\sim0.01~M_{\rm Moon}$, consistent with a lack 
of emission shortward of 25\,$\mu$m and a location $\lesssim$3\,AU from the 
star.  Future millimeter observations can test this assertion, as an inner dust 
belt should be readily separated from the central star at higher angular 
resolution.
\end{abstract}


\section{Introduction}

Debris disks are created by the collisional erosion of planetesimals, the 
building blocks of planetary systems.  These collisions continuously generate 
dust grains with a range of sizes that are detected with astronomical 
measurements from optical to radio wavelengths.  Resolved observations of 
nearby debris disks are instrumental in advancing our understanding of these 
systems.  At a distance of $9.91\pm0.10$~pc \citep{van07}, the M1 star AU~Mic 
hosts one of the closest and best studied debris disks.  The detection of 
submillimeter emission \citep{liu04b} from this $\sim$10~Myr-old system in the 
$\beta$~Pic moving group \citep{zuc01} was followed quickly by the discovery of 
an edge-on disk seen in scattered starlight \citep{kal04}.  Subsequent work has 
characterized the scattered light in great detail, exploiting its proximity to 
constrain its radial and vertical structure 
\citep{liu04a,kri05,met05,gra07,fit07}. 

Observations of dust emission at (sub)millimeter wavelengths provide important, 
complementary information about debris disk structures.  Unlike the small 
grains probed at optical and near-infrared wavelengths that react strongly to 
stellar radiation and wind forces, the large grains that dominate the 
millimeter-wave emission have dynamics more like the parent planetesimals.  As 
a result, long-wavelength images trace best the location and distribution of 
the larger colliding bodies \citep{wya06}, and potentially also the signatures 
of planets that interact with them \citep{ert12}.  These size-dependent dust 
dynamics manifest beautifully in the edge-on AU~Mic disk.  Resolved 
millimeter-wave observations show an emission belt within the extended optical 
disk that peaks near a radius of $35$~AU, where the midplane scattered light 
profile steepens dramatically \citep{wil12}.  These features are elegantly 
explained by the presence of a ``birth ring'' of planetesimals at that 
location, where small grains released in a collisional cascade are launched 
into an extended halo \citep{str06,aug06}.  

With the advent of the Atacama Large Millimeter Array (ALMA), the millimeter 
emission in nearby debris disks can be imaged in much greater detail 
\citep[e.g.,][]{bol12}.  In this {\em Letter}, we present new, sub-arcsecond 
resolution ALMA Cycle~0 observations of AU~Mic at $\lambda = 1.3$~mm.  The ALMA 
data provide substantially improved constraints on the locations of colliding 
planetesimals in the AU~Mic disk and help shed light on the processes that may 
be shaping the planetesimal distribution.  They also reveal a previously 
unknown, centrally located emission feature.

\section{Observations} \label{sec:obs}

AU~Mic was observed by ALMA with its Band~6 receivers over four 2\,hour-long 
``scheduling blocks" (SBs) in 2012 Apr and Jun.  Table~\ref{tab:obs} summarizes 
the observations.  The 16-20 operational 12-m antennas were arranged to span 
baseline lengths of 21--402\,m (corresponding to a maximum resolution of 
$\sim$0\farcs6).  The correlator was configured to optimize continuum 
sensitivity, processing two polarizations in four 2\,GHz-wide basebands, each 
with 128 spectral channels, centered at 226, 228, 242, and 244\,GHz.  In each 
SB, we interleaved observations of AU~Mic (pointing center $\alpha=20^{\rm 
h}45^{\rm m}09\fs34$, $\delta=-31\degr20\arcmin24\farcs09$, J2000, within 
1\arcsec\ of the star position at all epochs) with the nearby quasar 
J2101$-$295.

The data from each SB were calibrated independently within the {\tt CASA} 
software package.  After applying system temperature measurements and phase 
corrections from the water vapor radiometers, the data were flagged and 
averaged into 6.048\,s integrations.  A calibration of the spectral response of 
the system was determined from observations of J1924$-$292, and complex gain 
variations induced by atmospheric and instrumental effects were corrected using 
observations of J2101$-$295.  The absolute flux calibration was derived from 
observations of Neptune: a mean calibration was applied to all basebands, with 
a systematic uncertainty of $\sim$10\%\ (see \S 3.3).  To generate an image at 
the mean frequency, 235\,GHz (1.28\,mm), we Fourier inverted the calibrated 
visibilities with natural weighting and performed a multi-frequency synthesis 
deconvolution with the {\tt CLEAN} algorithm.  The visibilities were further 
reduced by spectrally averaging over the central 112 channels in each baseband 
and re-weighted by the observed scatter.

\section{Results and Analysis} \label{sec:results}

\subsection{Image of 1.3\,mm Dust Continuum Emission} \label{sec:continuum}

\begin{figure}[t]
\epsscale{0.7}
\plotone{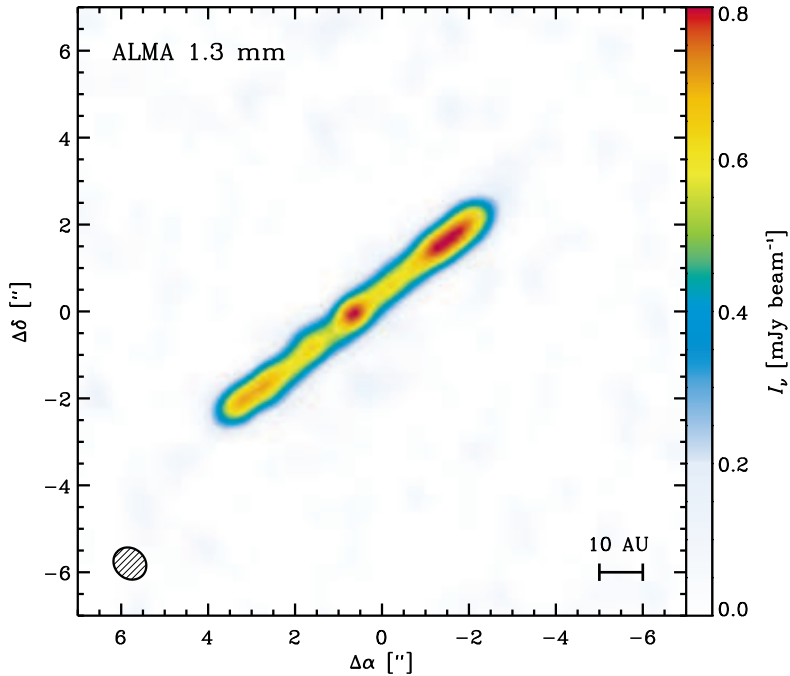}
\figcaption{
ALMA image of the 1.3\,mm continuum emission from AU~Mic.  The ellipse in the 
lower left corner represents the $0\farcs80 \times 0\farcs69$ ($8\times7$~AU) 
synthesized beam.  \label{fig:image}}
\end{figure}

Figure \ref{fig:image} shows an image of the $\lambda= 1.3$~mm emission from 
SB-4 (with the most antennas and best weather conditions), with synthesized 
beam $0\farcs80 \times 0\farcs69$ ($8\times7$~AU), p.a.~49\degr, and rms of 
30\,$\mu$Jy~beam$^{-1}$.  An image constructed from all 4 SBs is consistent but 
noisier, which we attribute to systematic calibration issues resulting from the 
poorer weather conditions of the earlier observations.  The emission is 
confined to a narrow band with aspect ratio $>$10:1, with an orientation 
consistent with the scattered light disk.  The emission is not resolved in the 
direction perpendicular to the elongation.  There are clear peaks near both 
extrema and in the middle of the structure (detected at all four epochs).  The 
emission is marginally brighter at the northwest end than the southeast end, 
and shows small undulations along its length, though none of these variations 
are significant.  We interpret the observed structure as a superposition of two 
components: (1) the nearly edge-on dust belt with limb-brightened ansae, and 
(2) a new, distinct, and compact feature located at the center of the belt.

\subsection{Modeling Formalism} \label{sec:model_formalism}

Building on the phenomenological methodology of \citet{wil11,wil12} to analyze 
resolved millimeter emission from debris disks, we construct a parametric model 
to quantify the observed properties of the AU~Mic emission.  We consider two 
model components: a vertically thin, axisymmetric ``outer" belt, and an 
additional source to account for the central peak.  The belt component is 
informed by models of the scattered light that show the disk midplane within 
50\,AU is remarkably straight, $\lesssim$0\fdg5 from edge-on, and thin (FWHM 
$\sim$0\farcs3).  We assume the belt is viewed at an inclination of 89\fdg5.  
The belt is characterized as an annulus with (unprojected) radial intensity 
$I_{\nu}(r)\propto r^x$ for $r_i < r < r_o$, with a normalization defined by 
$F_{\rm belt} = \int I_{\nu}\, d\Omega$, a center determined by offsets 
(relative to the pointing center) \{$\Delta \alpha$, $\Delta \delta$\}, and an 
orientation described by a position angle (PA).  We treat the central component 
as a circular Gaussian with mean $\Delta r_{\rm cen}$, variance 
$\sigma^{2}_{\rm cen}$ (half width at half maximum $\mathcal{R}_{\rm cen} = 
\sqrt{2 \ln{2}} \,\, \sigma_{\rm cen}$), and flux density $F_{\rm cen}$.  The 
mean $\Delta r_{\rm cen}$ is defined as a radial shift from the belt center 
{\it in the plane of the belt}.  We also include power-law spectral scalings 
between the 4 basebands for each component, denoted $\alpha_{\rm belt}$ and 
$\alpha_{\rm cen}$, where $F_{\nu} \propto \nu^{\alpha}$.  

For a given parameter set, we compute four synthetic visibility sets sampled at 
the same spatial frequencies observed by ALMA, corresponding to the spectrally 
averaged basebands (at 226, 228, 242, and 244\,GHz).  By fitting the visibility 
data directly, we are not sensitive to the non-linear effects of deconvolution, 
and take advantage of the full range of available spatial frequencies.  The fit 
quality is quantified by a likelihood metric, $\mathcal{L}$, determined from 
the $\chi^2$ values summed over the real and imaginary components at all 
spatial frequencies ($ \ln{\mathcal{L}} = -\chi^2 /2$).  A Monte Carlo Markov 
Chain (MCMC) approach was utilized to characterize the multi-dimensional 
parameter space of this model and determine the posterior probability 
distribution functions for each parameter.  We used the affine-invariant 
ensemble sampler proposed by \citet{goo10}, in a locally-modified version of 
the parallelized implementation described by \citet{for12}, to compute 
likelihood values for $\sim$10$^6$ MCMC trials.  Uniform priors were assumed 
for all parameters, with bounds imposed to ensure that the model was 
well-defined: \{$F_{\rm belt}$, $F_{\rm cen}$, $\sigma^2_{\rm cen}\} \ge 0$, 
and $0 \le r_i < r_o$.

\subsection{Results of Model Fits} \label{sec:modelfits}

\begin{figure}[t]
\begin{center}
\epsscale{1.05}
\plotone{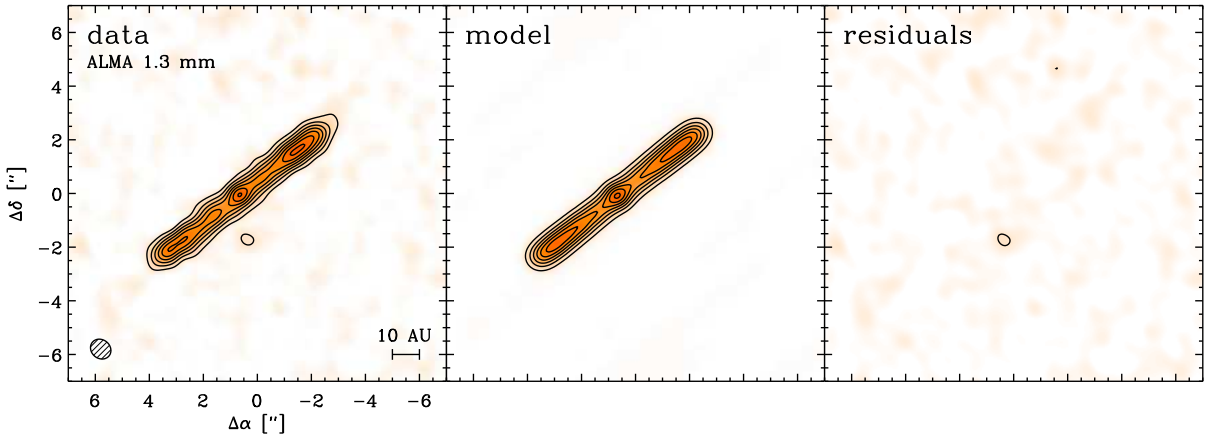}
\figcaption{
{\em (left)} The observed 1.3\,mm emission from AU~Mic, {\em (center)} the 
best-fit model (see \S\ref{sec:modelfits}), and {\em (right)} the imaged 
residuals.  Contours are drawn at 4\,$\sigma$ (120\,$\mu$Jy beam$^{-1}$) 
intervals.  \label{fig:model}}
\end{center}
\end{figure}

The best-fit parameter values and their 68\%\ uncertainties determined from 
the marginalized posterior probability distributions are listed in Table 
\ref{tab:model}.  The data and best-fit model are compared in the image plane 
in Figure~\ref{fig:model}; there are no significant residuals.  The best-fit 
model has a reduced $\chi^2 = 1.37$ (905,920 independent datapoints, 12 free 
parameters).  The modeling procedure was performed on each SB individually
and the full dataset (all 4 SBs together).  The results were entirely 
consistent, although the parameter uncertainties were notably smaller from the 
superior SB-4 dataset alone, and we focus on those results.

Most parameters are determined with high precision.  We find good agreement 
of the outer belt parameters \{$F_{\rm belt}$, $r_i$, $r_o$\} with the less 
well-constrained fits of \citet{wil12}, and on the disk PA from measurements of 
scattered starlight \citep[e.g.,][]{kri05}.  We measure a flat spectrum for the 
outer belt ($\alpha_{\rm belt} \approx 0$) across the 4 basebands, which 
corresponds to the {\em difference} between the spectral slopes of AU~Mic and 
Neptune ($\alpha_{\rm Neptune} \approx 2.1$), consistent with data from 
350\,$\mu$m to 1.3\,mm \citep{wil12}. 

The central emission peak is detected with high confidence at $F_{\rm cen} = 
320$\,$\mu$Jy ($>$10\,$\sigma$ brighter than the outer belt at that location).  
It is unresolved, with $\mathcal{R_{\rm cen}} \le 3.0$\,AU (3\,$\sigma$), and 
positionally coincident with the outer belt center: $\Delta r_{\rm cen} \le 
1.9$\,AU (3\,$\sigma$).  Regarding the outer belt, the most notable result is 
that the models strongly favor rising emission profiles with large, positive 
gradients: $x \approx 2.3\pm0.3$.  Models with the standard assumption of $x < 
0$ produce significant residuals, under-predicting the intensities at 
$\pm$1-2\arcsec\ from the belt center.  Because of the steep increase in the 
emission profile, there is only a weak constraint on the inner edge of the 
outer belt.  The best-fit $r_i$ deviates from 0 at the $\sim$2\,$\sigma$ level: 
the 3\,$\sigma$ limit is $r_i \le 21$\,AU.

\section{Discussion} \label{sec:discussion}

We have presented new, sub-arcsecond resolution ALMA observations of 1.3\,mm 
emission from the AU~Mic debris disk and analyzed the data with a simple 
parametric model.  This emission is resolved into two distinct components: 
(1) an edge-on outer belt with an emission profile that {\it rises} with radius out to 40~AU, and (2) an unresolved peak at the center of the outer belt.  
This distribution is more complex than the single, narrow ring often assumed 
for debris disks. However, it has some similarities to other nearby 
resolved systems, like $\epsilon$~Eri \citep{bac09} or HR~8799 \citep{su09}, 
that show an inner component inferred from excess infrared emission, separate 
from an extended and colder outer belt.  

\subsection{The Central Emission Peak} \label{sec:inner_belt}

The stellar photosphere is much fainter than the central peak noted in Figure 
\ref{fig:image}.  A NextGen stellar model \citep{hau99} with $T_{\rm eff} = 
3720$\,K, $L_{\ast} = 0.11$\,L$_{\odot}$, and $M_{\ast} = 0.6$\,M$_{\odot}$ 
\citep[e.g.,][]{met05,che05} that matches the AU~Mic photometry from 
0.4--25\,$\mu$m contributes only $F_{\ast} = 52$\,$\mu$Jy at 1.3\,mm, 
$\sim$6$\times$ fainter than observed.  However, AU~Mic is an active star that 
exhibits radio-wave bursts.  In quiescence, observations find $<$120\,~$\mu$Jy 
at 3.6\,cm \citep{whi94}, and the contribution at 1.3\,mm from hot coronal 
plasma seen in X-rays is unlikely to be significant \citep[though better 
spectral constraints are desirable, see][]{let00}.  Flares are detected from 
AU~Mic at $\sim$200-1200\,$\mu$Jy at 6\,cm \citep{bow09}, but this non-thermal 
emission is much weaker at 1.3\,mm.  While the unknown variability makes any 
extrapolation to 1.3\,mm problematic, the temporal properties of the ALMA 
emission provide additional information.  Radio-wave flares have fast decay 
times, of order an hour \citep{kun87}; but, the mm-wave peak persists at 
a consistent intensity in all four ALMA observations, within uncertainties
that are typically $2-3\times$ larger than for SB-4, spanning timescales 
from 1 hour (within SB-4) to 2 months (SB-1 to SB-4).  Unfortunately, the 
spectral index ($\alpha_{\rm cen}$) constraints are not good enough to be 
diagnostic.  We suspect that stellar 
emission is too weak and too ephemeral to be responsible for the 1.3\,mm peak, 
but the available data does not allow for a firm determination of its 
contribution. 

Alternatively, the central emission peak could be produced by dust in a 
distinct (unresolved) planetesimal belt located close to the star.  In \S 3.3, 
we constrained the extent of this peak to $\mathcal{R_{\rm cen}} \le 3$\,AU 
(3\,$\sigma$), inside the inner working angle ($0\farcs8 \approx 8$\,AU) of all 
previous high resolution imaging of scattered light \citep{kri05,fit07}.  Rough 
models of the spectral energy distribution (SED) from the ALMA central peak 
can help assess the feasibility that it originates in an inner dust belt.  
In this context, the most salient feature of the AU~Mic SED is the absence of 
emission excess at $\lambda \le 25$\,$\mu$m \citep[e.g.,][]{liu04b,che05}.  
We assume the central peak represents the combined emission from the star and 
dust, such that $F_{\rm dust} = F_{\rm cen} - F_{\ast} \approx 0.25$\,mJy at 
1.3\,mm.  Optically thin dust emission at a temperature, $T$, has $F_{\rm dust} 
\approx \kappa_{\nu} B_{\nu}(T)  M_{\rm dust} / d^2$, where $\kappa_{\nu}$ is 
the opacity spectrum, $B_{\nu}$ the Planck function, $M_{\rm dust}$ the mass, 
and $d = 9.91$\,pc.  For a given dust population characterized by 
$\kappa_{\nu}$, we computed the maximum $T$ (and minimum $M_{\rm dust}$) 
consistent with both the observed millimeter flux density and the infrared 
SED.  We calculated various $\kappa_{\nu}$ for dust with the \citet{wei01} 
``astrosilicate'' composition and a power-law size distribution $n(a) \propto 
a^{-3.5}$ between $a_{\rm min} = 0.2$\,$\mu$m \citep[the blow-out 
size;][]{str06} and $a_{\rm max}$ values from 1\,$\mu$m to 1\,cm.  
For $a_{\rm max} \le 100$ $\mu$m, models of the central peak over-predict 
the observed 60-70 $\mu$m emission if $T > 35$ K.  However, larger grains 
with $a_{\rm max} \ge 1$~mm at temperatures up to $T \approx 75$\,K 
can be accommodated without producing an excess at $\lambda \le 25$ $\mu$m.  
These maximum $T$ values are comparable to the expected dust temperatures a few 
AU from the star, compatible with the emission size constraints 
($\mathcal{R_{\rm cen}}$).  The corresponding minimum $M_{\rm dust}$ is 
$\sim$$9 \times 10^{23}$\,g, about 1\%\ of the lunar mass.  These calculations 
show that the central emission peak is consistent with a cool dust belt located 
$\lesssim$3\,AU from the central star, with a total mass comparable to the 
asteroid belt in our Solar System.  If this interpretation is correct, then 
ALMA observations at higher resolution can determine its properties. 
Interestingly, the temperature of this putative inner belt is colder than 
the $\sim190$\,K found to apply systematically to inner belts around F5-K0 
stars by \citet{mor11}.

\subsection{The Outer Dust Belt} \label{sec:outer_belt}

Our modeling of the ALMA data locates the far edge of the outer emission belt 
with high precision, $r_o = 40$\,AU, which matches closely the outer edge 
of the hypothesized ``birth ring'' of colliding planetesimals.  This analysis 
does not define the shape of the edge below the $\sim6$\,AU resolution limit, 
but the truncation is reminiscent of the outer edge of the classical Kuiper 
Belt \citep[$47\pm1$\,AU;][]{tru01}.  The origins of such sharp edges remain 
unclear: they could be from dynamical interactions \citep{ida00,bol12}, or they 
may simply represent the initial conditions, where planetesimal formation was 
efficient and successful in the primordial disk.  Adopting the opacity used in 
\S 4.1 ($\kappa_{\nu}$ = 2.7\,cm$^2$\,g$^{-1}$), and assuming $T \approx 25$\,K 
(for 35-45\,AU), the dust mass of this outer belt is $7 \times 10^{25}$\,g 
\citep[consistent with previous estimates;][]{liu04b}, $\sim$100$\times$ more 
massive than the hypothesized inner belt; the Kuiper Belt and asteroid belt 
have a similar mass ratio.  

The mm-wave emission morphologies of cold belts of dusty debris reflect the 
dynamical processes that shape the underlying planetesimal distributions.  For 
AU~Mic, our modeling suggests that its outer emission belt can be described 
by an increasing emission profile with a positive radial power-law index $x 
\approx 2.3\pm0.3$.  If we assume the emitting dust is in radiative equilibrium 
with a temperature profile $T \propto r^{-0.5}$, this implies a rising surface 
density profile, $\Sigma \propto r^{2.8}$, strongly peaked near 40\,AU.  A 
broad parent body ring with constant surface density would produce a radial 
intensity profile with $x\approx -0.5$, a value ruled out with high confidence 
($>5\sigma$).  A rising behavior is predicted for ``self-stirred" disks with 
ongoing planet formation \citep{ken02}; in particular, the models of 
\citet{ken10} suggest $\Sigma \propto r^{7/3}$.  However, the timescale 
required to assemble Pluto-sized bodies at $\sim$40\,AU to initiate a 
collisional cascade around a low-mass star like AU~Mic is much longer than its 
$\sim$10\,Myr age \citep{ken08}.  Moreover, this scenario does not naturally 
accommodate the presence of a separate, interior planetesimal belt.  Of course, 
the still modest resolution of the data is compatible with more complex 
scenarios, such as multiple closely-spaced belts of different brightnesses that 
mimic a smooth gradient.  Scattered light observations of the AU~Mic disk show 
asymmetries on both large and small scales, with several peaks and depressions 
projected against the broad ansae in Figure \ref{fig:image}, at radii beyond 
the millimeter undulations \citep[features A-E; see][]{fit07}.  With such a 
steep emission gradient in this outer belt, the data do not strongly constrain 
its width, or the location of its inner edge.  Our modeling indicates 
substantial emission from mm-sized grains interior to 40\,AU, in the 
$\sim$20-40\,AU zone inferred to be highly depleted of $\mu$m-sized grains from 
polarized scattered light \citep{gra07}.

The ALMA data show no clear evidence for asymmetries or substructure that would 
signal planet-disk interactions.  The hints of modulating millimeter brightness 
along the belt in Figure~\ref{fig:image} are insignificant in the residuals 
from subtracting a symmetric parametric model (see Figure \ref{fig:model}).  
This rules out substructure brighter than 90\,$\mu$Jy beam$^{-1}$ 
(3\,$\sigma$), corresponding to dust clumps $\gtrsim 1$\%\ of the lunar mass 
(for the dust properties adopted above).  Those limits argue against 
over-densities of dust-producing planetesimals trapped in mean motion 
resonances \citep{kuc03}, as might arise from the outward migration of planets 
\citep{wya03}.  Given the young age of the system, the broad and smooth 
character of the outer belt in the AU~Mic disk may resemble the Kuiper Belt 
prior to the epoch of Neptune's migration \citep{mal95}.  It is interesting 
that none of the claims of millimeter emission clumps in debris disks have 
survived scrutiny at higher sensitivity \citep{pie11,hug11,hug12}.  It may be 
that any such features are effectively erased by collisions \citep{kuc10}.  We 
also find no significant centroid offset between the outer belt and central 
peak, as might result from the secular perturbations of a planet in an 
eccentric orbit \citep{wya99}.  The limit on the displacement, $\Delta r_{\rm 
cen} < 1.9$\,AU (3\,$\sigma$), corresponds approximately to a limit on $ae$, 
where $a$ is the semi-major axis and $e$ is the eccentricity.  This limit can 
still accommodate a wide-orbit planet with modest eccentricity, similar to 
Uranus.  Such a planet could be responsible for stirring the disk to 40\,AU in 
$\sim$10~Myr \citep[e.g. for $a=30$\,AU and $e=0.05$, see eqn. 15 of][]{mus09}.
Limits from high contrast direct imaging admit Saturn-mass planets at these 
separations \citep{del12}.

\subsection{Concluding Remarks}

The basic architecture of the AU~Mic debris disk appears remarkably similar to 
the Solar System, with a potential analog to the asteroid belt at a few AU, and 
a colder, more massive, and apparently truncated counterpart of the Kuiper Belt 
extending to 40\,AU.  Future observations are needed to determine if stellar 
processes could be responsible for emission attributed to the asteroid belt, 
and to determine if the Solar System analogy extends to include a planetary 
system like our own.

\acknowledgments{
M.A.M.~thanks NRAO for Student Observing Support funds.  A.M.H.~is supported 
by a fellowship from the Miller Institute for Basic Research in Science.  
M.B. is funded through a Space Science Enhancement Program grant from the 
Canadian Space Agency and an NSERC Discovery Accelerator Supplement.
E.C. acknowledges NSF grant AST-0909210.  
P.K. and J.R.G. acknowledge support from NSF Award 0909188 and NASA Award
NNX11AD21G
This paper makes use of the following ALMA data: ADS/JAO.ALMA\#2011.0.00142.S.  
ALMA is a partnership of ESO (representing its member states), NSF (USA) and 
NINS (Japan), together with NRC (Canada) and NSC and ASIAA (Taiwan), in 
cooperation with the Republic of Chile. The Joint ALMA Observatory is operated 
by ESO, AUI/NRAO and NAOJ.  The National Radio Astronomy Observatory is a 
facility of the National Science Foundation operated under cooperative agreement 
by Associated Universities, Inc.
}

{\it Facility:} \facility{ALMA}

\clearpage

\begin{deluxetable}{cccc}
\tablecaption{ALMA Cycle 0 Observations of AU Mic}
\tablewidth{0pt}
\tablehead{
\colhead{ID} & \colhead{Date (UT)} &
\colhead{Antennas} & \colhead{PWV (mm)}
}
\startdata
SB-1 & 2012 Apr 23  07:30 -- 09:26 & 17 & 1.7 \\
SB-2 & 2012 Apr 23  09:39 -- 11:03 & 16 & 1.7 \\
SB-3 & 2012 Apr 24  09:09 -- 11:19 & 18 & 3.0 \\
SB-4 & 2012 Jun 16  05:48 -- 08:02 & 20 & 0.7 \\
\enddata
\label{tab:obs}
\end{deluxetable}

\begin{deluxetable}{clrc}
\tablecaption{Model Parameters}
\tablewidth{0pt}
\tablehead{
\colhead{Parameter} & \colhead{Description} &
\colhead{Best-Fit} & \colhead{68\% Confidence Interval}
}
\startdata
$F_{\rm belt}$       & belt flux density (mJy)              &   7.14 & {+0.12}, {-0.25} \\
$x$                  & belt radial power law index          &   2.32 & {+0.21}, {-0.31} \\
$r_i$                & belt inner radius (AU)               &    8.8 & {+11.0}, {-1.0}   \\
$r_o$                & belt outer radius (AU)               &   40.3 & {+0.4}, {-0.4}   \\
PA                   & belt position angle (\degr)          & 128.41 & {+0.12}, {-0.13} \\
$\alpha_{\rm belt}$  & belt spectral index                  &  -0.15 & {+0.40}, {-0.58} \\
\hline
$F_{\rm cen}$        & Gaussian flux density (mJy)          &   0.32 & {+0.06}, {-0.06} \\
$\Delta r_{\rm cen}$ & Gaussian offset (AU)                 &   0.71 & {+0.35}, {-0.51} \\
$\sigma^2_{\rm cen}$ & Gaussian variance (AU$^2$)           & $\le$5.9 & (3\,$\sigma$ limit) \\
$\alpha_{\rm cen}$   & Gaussian spectral index              &  -0.35 & {+2.1}, {-4.5} \\
\hline
$\Delta \alpha$      & R.A. offset of belt center (\arcsec) &   0.61 & {+0.02}, {-0.02} \\
$\Delta \delta $     & Dec. offset of belt center (\arcsec) &  -0.03 & {+0.02}, {-0.02} \\
\enddata
\label{tab:model}
\end{deluxetable}


\clearpage

\begin{thebibliography}{}

\bibitem[Augereau \& Beust(2006)]{aug06} Augereau, J.-C., \& Beust, H.\ 
  2006, \aap, 455, 987 

\bibitem[Backman et al.(2009)]{bac09} Backman, D., Marengo,
  M., Stapelfeldt, K., et al.\ 2009, ApJ, 690, 1522

\bibitem[Boley et al.(2012)]{bol12} Boley, A.~C., Payne, M.~J., Corder, S., 
  et al.\ 2012, \apjl, 750, L21 

\bibitem[Bower et al.(2009)]{bow09} Bower, G.~C., Bolatto, 
  A., Ford, E.~B., \& Kalas, P.\ 2009, \apj, 701, 1922 

\bibitem[Chen et al.(2005)]{che05} Chen, C.~H., Patten, 
  B.~M., Werner, M.~W., et al.\ 2005, \apj, 634, 1372 

\bibitem[Delorme et al.(2012)]{del12} Delorme, P., et al.\ 2012, \aap, 539, 72

\bibitem[Ertel et al.(2012)]{ert12} Ertel, S., Wolf, S., 
  \& Rodmann, J.\ 2012, \aap, 544, A61 

\bibitem[Fitzgerald et al.(2007)]{fit07} Fitzgerald, M.~P., Kalas, P.~G., 
  Duch{\^e}ne, G., Pinte, C., \& Graham, J.~R.\ 2007, \apj, 670, 536 

\bibitem[Foreman-Mackey et al.(2012)]{for12} Foreman-Mackey, D., Hogg, D. W., 
  Lang, D., \& Goodman, J. 2012, arXiv:1202.3665

\bibitem[Goodman \& Weare(2010)]{goo10} Goodman, J., \& Weare, J. 2010, 
  Comm. App. Math. Comp. Sci., 5, 65

\bibitem[Graham et al.(2007)]{gra07} Graham, J.~R., Kalas, P.~G., 
  \& Matthews, B.~C.\ 2007, \apj, 654, 595 

\bibitem[Hauschildt et al.(1999)]{hau99} Hauschildt, P. H., Allard,
  F., \& Baron, E. 1999, \apj, 512, 377

\bibitem[Hughes et al.(2011)]{hug11} Hughes, A.~M., Wilner, 
  D.~J., Andrews, S.~M., et al.\ 2011, \apj, 740, 38 

\bibitem[Hughes et al.(2012)]{hug12} Hughes, A.~M., Wilner,
  D.~J., Mason, B., et al.\ 2012, ApJ, 750, 82

\bibitem[Ida et al.(2000)]{ida00} Ida, S., Larwood, J., \& Burkert, A. 
  2000, \apj, 528, 351

\bibitem[Kalas et al.(2004)]{kal04} Kalas, P., Liu, M.~C., 
  \& Matthews, B.~C.\ 2004, Science, 303, 1990 

\bibitem[Kennedy \& Wyatt(2010)]{ken10} Kennedy, G.~M., \& Wyatt, M.~C.\ 
  2010, \mnras, 405, 1253 

\bibitem[Kenyon \& Bromley(2002)]{ken02} Kenyon, S.~J., \& Bromley, B.~C.\ 
  2002, \apjl, 577, L35 

\bibitem[Kenyon \& Bromley(2008)]{ken08} Kenyon, S.~J., \& Bromley, B.~C.\ 
  2008, \apjs, 179, 451 

\bibitem[Krist et al.(2005)]{kri05} Krist, J.~E., Ardila, 
D.~R., Golimowski, D.~A., et al.\ 2005, \aj, 129, 1008 

\bibitem[Kuchner \& Holman(2003)]{kuc03} Kuchner, M.~J., \& Holman, M.~J.\ 
  2003, \apj, 588, 1110 

\bibitem[Kuchner \& Stark(2010)]{kuc10} Kuchner, M.~J., \& Stark, C.~C.\ 
  2010, \aj, 140, 1007 

\bibitem[Kundu et al.(1987)]{kun87} Kundu, M.~R., Jackson, 
  P.~D., White, S.~M., \& Melozzi, M.\ 1987, \apj, 312, 822 

\bibitem[Leto et al.(2000)]{let00} Leto, G., Pagano, I., Linsky, J.~L.,
  Rodon{\`o}, M., \& Umana, G.\ 2000, \aap, 359, 1035

\bibitem[Liu(2004)]{liu04a} Liu, M.~C.\ 2004, Science, 305, 1442 

\bibitem[Liu et al.(2004)]{liu04b} Liu, M.~C., Matthews, 
  B.~C., Williams, J.~P., \& Kalas, P.~G.\ 2004, \apj, 608, 526 

\bibitem[Malhotra(1995)]{mal95} Malhotra, R.\ 1995, \aj, 110, 420 

\bibitem[Metchev et al.(2005)]{met05} Metchev, S.~A., Eisner, 
  J.~A., Hillenbrand, L.~A., \& Wolf, S.\ 2005, \apj, 622, 451 

\bibitem[Morales et al.(2011)]{mor11} Morales, F.~Y., Rieke, 
  G.~H., Werner, M.~W., et al.\ 2011, \apjl, 730, L29 

\bibitem[Mustill \& Wyatt(2009)]{mus09} Mustill, A.~J., \& Wyatt, M.~C.\ 2009, 
  \mnras, 399, 1403 

\bibitem[Pi{\'e}tu et al.(2011)]{pie11} Pi{\'e}tu, V., di Folco, E.,
  Guilloteau, S., Gueth, F., \& Cox, P.\ 2011, A\&A, 531, L2

\bibitem[Su et al.(2009)]{su09} Su, K.~Y.~L., Rieke, G.~H., 
  Stapelfeldt, K.~R., et al.\ 2009, \apj, 705, 314 

\bibitem[Strubbe \& Chiang(2006)]{str06} Strubbe, L.~E., \& Chiang, E.~I.\ 
  2006, \apj, 648, 652 

\bibitem[Trujillo \& Brown(2001)]{tru01} Trujillo, C.~A., \& Brown, M.~E.\ 
  2001, \apjl, 554, L95 

\bibitem[van Leeuwen(2007)]{van07} van Leeuwen, F.\ 2007, \aap, 474, 653 

\bibitem[Weingartner \& Draine(2001)]{wei01} Weingartner, J. C.,
  \& Draine, B. T. 2001, \apj, 548, 296

\bibitem[White et al.(1994)]{whi94} White, S.~M., Lim, J.,
  \& Kundu, M.~R.\ 1994, \apj, 422, 293

\bibitem[Wilner et al.(2011)]{wil11} Wilner, D.~J., Andrews, 
  S.~M., \& Hughes, A.~M.\ 2011, \apjl, 727, L42 

\bibitem[Wilner et al.(2012)]{wil12} Wilner, D.~J., Andrews, 
  S.~M., MacGregor, M.~A., \& Hughes, A.~M.\ 2012, \apjl, 749, L27 

\bibitem[Wyatt et al.(1999)]{wya99} Wyatt, M.~C., Dermott, 
  S.~F., Telesco, C.~M., et al.\ 1999, \apj, 527, 918 

\bibitem[Wyatt(2003)]{wya03} Wyatt, M.~C.\ 2003, \apj, 598, 
1321 

\bibitem[Wyatt(2006)]{wya06} Wyatt, M.~C.\ 2006, \apj, 639, 1153 

\bibitem[Zuckerman et al.(2001)]{zuc01} Zuckerman, B., Song, 
I., Bessell, M.~S., \& Webb, R.~A.\ 2001, \apjl, 562, L87 


\end{thebibliography}
\end{document}